\input harvmac
\def\ev#1{\langle#1\rangle}
\input amssym
\input epsf
\let\includefigures=\iftrue
%
% the following is to use blackboard bold fonts --
%\let\useblackboard=\iftrue
%
% activate this if you don't have them.
%\let\useblackboard=\iffalse
%
% You might also need to remove this line.
\newfam\black
\noblackbox
\includefigures
\message{If you do not have epsf.tex (to include figures),}
\message{change the option at the top of the tex file.}
\def\figin{\epsfcheck\figin}\def\figins{\epsfcheck\figins}
\def\epsfcheck{\ifx\epsfbox\UnDeFiNeD
\message{(NO epsf.tex, FIGURES WILL BE IGNORED)}
\gdef\figin##1{\vskip2in}\gdef\figins##1{\hskip.5in}% blank space instead
\else\message{(FIGURES WILL BE INCLUDED)}%
\gdef\figin##1{##1}\gdef\figins##1{##1}\fi}
\def\DefWarn#1{}

\def\figinsert{\goodbreak\midinsert}
\def\ifig#1#2#3{\DefWarn#1\xdef#1{fig.~\the\figno}
\writedef{#1\leftbracket fig.\noexpand~\the\figno}%
\figinsert\figin{\centerline{#3}}\medskip\centerline{\vbox{\baselineskip12pt
\advance\hsize by -1truein\noindent\footnotefont{\bf
Fig.~\the\figno:} #2}}
\bigskip\endinsert\global\advance\figno by1}
%%%
\else
\def\ifig#1#2#3{\xdef#1{fig.~\the\figno}
\writedef{#1\leftbracket fig.\noexpand~\the\figno}%
%\figinsert\figin{\centerline{#3}}\medskip\centerline{\vbox{\baselineskip12pt
%\advance\hsize by -1truein\noindent\footnotefont{\bf Fig.~\the\figno:} #2}}
%\bigskip\endinsert
\global\advance\figno by1} \fi

%%%%%%% References %%%%%%%

\def\det{{\rm det}}

%% MACROS

\def\IL{\relax{\rm I\kern-.18em L}}
\def\IH{\relax{\rm I\kern-.18em H}}
\def\IR{\relax{\rm I\kern-.18em R}}
\def\IC{\relax\hbox{$\inbar\kern-.3em{\rm C}$}}
\def\IZ{\relax\ifmmode\mathchoice
{\hbox{\cmss Z\kern-.4em Z}}{\hbox{\cmss Z\kern-.4em Z}}
{\lower.9pt\hbox{\cmsss Z\kern-.4em Z}} {\lower1.2pt\hbox{\cmsss
Z\kern-.4em Z}}\else{\cmss Z\kern-.4em Z}\fi}
\def\CM {{\cal M}}

%% MORE MACROS
\def\CM {{\cal M}}

\def\det{{\rm det}}
\def\Tr{{\rm Tr}}

\font\manual=manfnt \def\dbend{\lower3.5pt\hbox{\manual\char127}}

\def\IZ{\relax\ifmmode\mathchoice
{\hbox{\cmss Z\kern-.4em Z}}{\hbox{\cmss Z\kern-.4em Z}}
{\lower.9pt\hbox{\cmsss Z\kern-.4em Z}} {\lower1.2pt\hbox{\cmsss
Z\kern-.4em Z}}\else{\cmss Z\kern-.4em Z}\fi}

\def\bar{\overline}

\def\rt2{\sqrt{2}}
\def\irt2{{1\over\sqrt{2}}}

%  \slashchar puts a slash through a character to represent contraction
%  with Dirac matrices. Use \not instead for negation of relations, and use
%  \hbar for hbar.
\def\slashchar#1{\setbox0=\hbox{$#1$}           % set a box for #1
   \dimen0=\wd0                                 % and get its size
   \setbox1=\hbox{/} \dimen1=\wd1               % get size of /
   \ifdim\dimen0>\dimen1                        % #1 is bigger
      \rlap{\hbox to \dimen0{\hfil/\hfil}}      % so center / in box
      #1                                        % and print #1
   \else                                        % / is bigger
      \rlap{\hbox to \dimen1{\hfil$#1$\hfil}}   % so center #1
      /                                         % and print /
   \fi}

%\AffleckRR
\lref\AffleckRR{
  I.~Affleck, M.~Dine and N.~Seiberg,
  ``Supersymmetry Breaking By Instantons,''
  Phys.\ Rev.\ Lett.\  {\bf 51}, 1026 (1983).
  %%CITATION = PRLTA,51,1026;%%
}
%\AffleckMK
\lref\AffleckMK{
  I.~Affleck, M.~Dine and N.~Seiberg,
  ``Dynamical Supersymmetry Breaking In Supersymmetric QCD,''
  Nucl.\ Phys.\ B {\bf 241}, 493 (1984).
  %%CITATION = NUPHA,B241,493;%%
}
%\AffleckUZ
\lref\AffleckUZ{
  I.~Affleck, M.~Dine and N.~Seiberg,
  ``Calculable Nonperturbative Supersymmetry Breaking,''
  Phys.\ Rev.\ Lett.\  {\bf 52}, 1677 (1984).
  %%CITATION = PRLTA,52,1677;%%
}
%\AffleckXZ
\lref\AffleckXZ{
  I.~Affleck, M.~Dine and N.~Seiberg,
  ``Dynamical Supersymmetry Breaking In Four-Dimensions And Its
  Phenomenological Implications,''
  Nucl.\ Phys.\ B {\bf 256}, 557 (1985).
  %%CITATION = NUPHA,B256,557;%%
}
%\CachazoSG
\lref\CachazoSG{
  F.~Cachazo, B.~Fiol, K.~A.~Intriligator, S.~Katz and C.~Vafa,
  ``A geometric unification of dualities,''
  Nucl.\ Phys.\ B {\bf 628}, 3 (2002)
  [arXiv:hep-th/0110028].
  %%CITATION = HEP-TH 0110028;%%
}
%\PoppitzVD
\lref\PoppitzVD{
  E.~Poppitz and S.~P.~Trivedi,
  ``Dynamical supersymmetry breaking,''
  Ann.\ Rev.\ Nucl.\ Part.\ Sci.\  {\bf 48}, 307 (1998)
  [arXiv:hep-th/9803107].
  %%CITATION = HEP-TH 9803107;%%
}
%\ShadmiJY
\lref\ShadmiJY{
  Y.~Shadmi and Y.~Shirman,
  ``Dynamical supersymmetry breaking,''
  Rev.\ Mod.\ Phys.\  {\bf 72}, 25 (2000)
  [arXiv:hep-th/9907225].
  %%CITATION = HEP-TH 9907225;%%
}
%\DouglasSW
\lref\DouglasSW{
  M.~R.~Douglas and G.~W.~Moore,
  ``D-branes, Quivers, and ALE Instantons,''
  arXiv:hep-th/9603167.
  %%CITATION = HEP-TH 9603167;%%
}
%\BerensteinXA
\nref\BerensteinXA{
  D.~Berenstein, C.~P.~Herzog, P.~Ouyang and S.~Pinansky,
  ``Supersymmetry breaking from a Calabi-Yau singularity,''
  JHEP {\bf 0509}, 084 (2005)
  [arXiv:hep-th/0505029].
  %%CITATION = HEP-TH 0505029;%%
}
%\FrancoZU
\nref\FrancoZU{
  S.~Franco, A.~Hanany, F.~Saad and A.~M.~Uranga,
  ``Fractional branes and dynamical supersymmetry breaking,''
  arXiv:hep-th/0505040.
  %%CITATION = HEP-TH 0505040;%%
}
%\BertoliniDI
\nref\BertoliniDI{
  M.~Bertolini, F.~Bigazzi and A.~L.~Cotrone,
  ``Supersymmetry breaking at the end of a cascade of Seiberg dualities,''
  Phys.\ Rev.\ D {\bf 72}, 061902 (2005)
  [arXiv:hep-th/0505055].
  %%CITATION = HEP-TH 0505055;%%
}
%\FengMI
\lref\FengMI{
  B.~Feng, A.~Hanany and Y.~H.~He,
  ``D-brane gauge theories from toric singularities and toric duality,''
  Nucl.\ Phys.\ B {\bf 595}, 165 (2001)
  [arXiv:hep-th/0003085].
  %%CITATION = HEP-TH 0003085;%%
}
\lref\DineXK{
  M.~Dine, N.~Seiberg and E.~Witten,
  ``Fayet-Iliopoulos Terms In String Theory,''
  Nucl.\ Phys.\ B {\bf 289}, 589 (1987).
  %%CITATION = NUPHA,B289,589;%%
}
%\HerzogTR
\lref\HerzogTR{
  C.~P.~Herzog, Q.~J.~Ejaz and I.~R.~Klebanov,
  ``Cascading RG flows from new Sasaki-Einstein manifolds,''
  JHEP {\bf 0502}, 009 (2005)
  [arXiv:hep-th/0412193].
  %%CITATION = HEP-TH 0412193;%%
}
%\SeibergBZ
\lref\SeibergBZ{
  N.~Seiberg,
  ``Exact results on the space of vacua of four-dimensional SUSY gauge
  theories,''
  Phys.\ Rev.\ D {\bf 49}, 6857 (1994)
  [arXiv:hep-th/9402044].
  %%CITATION = HEP-TH 9402044;%%
}
%\IntriligatorAU
\lref\IntriligatorAU{
  K.~A.~Intriligator and N.~Seiberg,
  ``Lectures on supersymmetric gauge theories and electric-magnetic  duality,''
  Nucl.\ Phys.\ Proc.\ Suppl.\  {\bf 45BC}, 1 (1996)
  [arXiv:hep-th/9509066].
  %%CITATION = HEP-TH 9509066;%%
}
%\DiaconescuPC
\lref\DiaconescuPC{
  D.~E.~Diaconescu, B.~Florea, S.~Kachru and P.~Svrcek,
  ``Gauge -- Mediated Supersymmetry Breaking in String Compactifications,''
  arXiv:hep-th/0512170.
  %%CITATION = HEP-TH 0512170;%%
}
\lref\Beasley{C.~Beasley, Senior Thesis, Duke University}

%\TerningTH
\lref\TerningTH{
  J.~Terning,
  ``Non-perturbative supersymmetry,''
  arXiv:hep-th/0306119.
  %%CITATION = HEP-TH 0306119;%%
}

%\AmatiFT
\lref\AmatiFT{
  D.~Amati, K.~Konishi, Y.~Meurice, G.~C.~Rossi and G.~Veneziano,
  ``Nonperturbative Aspects In Supersymmetric Gauge Theories,''
  Phys.\ Rept.\  {\bf 162}, 169 (1988).
  %%CITATION = PRPLC,162,169;%%
 }

\lref\SW{N. Seiberg and E.Witten, unpublished.}

\def\new{\refs{\BerensteinXA - \BertoliniDI}}

\newbox\tmpbox\setbox\tmpbox\hbox{\abstractfont }
\Title{\vbox{\baselineskip12pt \hbox{UCSD-PTH-05-17}}}
{\vbox{\centerline{The Runaway Quiver}}}
\smallskip
\centerline{Kenneth Intriligator$^{1,2}$ and Nathan Seiberg$^2$ }
\smallskip
\bigskip
\centerline{$^1${\it Department of Physics, University of
California, San Diego, La Jolla, CA 92093 USA}}
\medskip
\centerline{$^2${\it School of Natural Sciences, Institute for
Advanced Study, Princeton, NJ 08540 USA}}
\bigskip
\vskip 1cm

 \noindent
We point out that some recently proposed string theory
realizations of dynamical supersymmetry breaking actually do not
break supersymmetry in the usual desired sense. Instead, there is
a runaway potential, which slides down to a supersymmetric vacuum
at infinite expectation values for some fields. The runaway
direction is not on a separated branch; rather, it shows up as
a``tadpole" everywhere on the moduli space of field expectation
values.

\Date{December 2005}

\newsec{Introduction}

There has been some recent interest in finding string theory
realizations of dynamical supersymmetry breaking.  In this short
note, we point out that some recently proposed realizations
\new\  have runaway potentials, and therefore do not break
supersymmetry in the usual desired sense. Our analysis here is a
standard analysis of the field theory (for reviews see e.g.\
\refs{\IntriligatorAU,\TerningTH}), and some of our conclusions
may already be known to some experts. However, encouraged by other
experts, we will anyway present here our modest findings, in the
hope that some members of the community might find it useful.

Theories with unstable, runaway directions in field space are
generally not considered as viable models of supersymmetry
breaking\foot{They could, however, be useful for quintessence.}.
Runaway unstable modes can be regarded as a tadpoles, since they
lead to violation of the static equations of motion. In looking
for supersymmetry breaking (meta)stable ground states, one must
always be careful about this potential pitfall.

A simple example with runaway is $SU(2)$ gauge theory with $N_f=1$
flavor, with dynamical superpotential  for $\CM =Q\widetilde Q$
\AffleckMK\
 \eqn\wsuii{W={\Lambda ^5\over \CM }.}
For any finite $\ev{\CM}$, there is no groundstate satisfying the
static equations of motion. The potential sends
$\ev{\CM}\rightarrow \infty$, where asymptotically supersymmetry
is restored.

Supersymmetry breaking can also be phrased as an inability to
satisfy the chiral ring relations \refs{\SW,\BerensteinXA -
\BertoliniDI}, but one must still check whether there is a stable,
non-supersymmetric groundstate or a runaway direction. For the
above example, the chiral ring is generated by $\CM =Q\widetilde
Q$, and the glueball chiral superfield $S\sim \Tr W_\alpha
W^\alpha$.  The classical ring relations are $S\CM =0$ and $S^2=0$
\SW. They are deformed in the quantum theory to
 \eqn\wsuiicr{S\CM =\Lambda ^5, \qquad S^2=0.}
One way to see that is to follow the instanton calculations of
\AmatiFT. The quantum relations \wsuiicr\ are incompatible for any
finite $\ev{\CM}$, but are asymptotically satisfied along the
runaway direction $\ev{S}\rightarrow 0$, $\ev{\CM}\rightarrow
\infty$. One lesson from this example is that incompatible ring
relations do not necessarily mean that supersymmetry is broken.
Instead, there could be a runaway to a supersymmetric ground state
at infinity.

Calculable examples of dynamical supersymmetry breaking without
runaways were first presented in \refs{\AffleckUZ, \AffleckXZ}.  A
review and survey of other models can be found e.g. in
\refs{\TerningTH,\PoppitzVD, \ShadmiJY}.

It was recently suggested \new\ that a string theory realization
of dynamical supersymmetry breaking is obtained from IIB string
theory, with wrapped D5 branes (``fractional branes"), on a
Calabi-Yau space that is locally a complex cone over the surface
$F_1$ (a.k.a. $dP_1$).  See \DiaconescuPC\ for additional examples
and discussion.  The conformal gauge theory for D3 branes only on
the $dP_1$ geometry was given in \refs{\Beasley, \FengMI}, and the
non-conformal theory with added wrapped D5 branes in \CachazoSG.
The $dP_1$ geometry does not admit an analog of the conifold
deformation, which corresponds to the fact that, with wrapped D5
branes,  the IR limit of the gauge theory is not simply SUSY
Yang-Mills, with its gaugino condensation \CachazoSG. The
suggestion of \new\ is that the IR limit of the cascade exhibits
dynamical supersymmetry breaking.  This suggestion was entirely
based  on an analysis of the low-energy effective gauge theory.
The supergravity dual solution of \HerzogTR\ is singular in the
IR, and there is no presently known smoothed version to illuminate
the IR physics.

Here we point out that this quiver gauge theory has a runaway
unstable mode, everywhere on the moduli space (the runaway
direction is not on a separated branch).  There is no static
vacuum where supersymmetry is broken and the equations of motion
are satisfied.  Much as in \wsuii, the fields can always slide
down to lower energy values, and asymptotically supersymmetry is
restored (at infinite expectation values of some fields).

Our analysis is formulated in terms of the gauge theory. Perhaps
some string theory dynamics -- outside of the realm of the
low-energy field theory analysis --  somehow stabilizes the
runaway mode in a way that breaks supersymmetry.   In the context
of compactification on a compact Calabi-Yau, this question hinges
on whether and how a particular K\" ahler modulus is stabilized,
as was recently discussed in \DiaconescuPC.  This issue merits
further study.  Note that  the (singular) supergravity solution of
\HerzogTR\ is supersymmetric.

The runaway directions that we discuss were already noted in
the analysis of  \refs{\BerensteinXA, \FrancoZU, \DiaconescuPC}.
It was suggested \refs{\BerensteinXA, \DiaconescuPC}\  that the D-term potentials
of some $U(1)$ factors could be a cure. Here we stress that these
$U(1)$ factors are anomalous, and hence massive.  Therefore, their
D-term equations should not be imposed.  This is related
to comments about ``dynamical relaxation" that also already appear
in some subsections of \refs{\FrancoZU, \DiaconescuPC}.  
We feel that it is worth stressing the bottom
line: {\it anomalous $U(1)$ D-terms should not be imposed in the
low-energy theory, and they cannot prevent the runaway.}

Finally, we should stress that our field theory analysis relies on
a Lagrangian with canonical kinetic terms which are renormalized
by field theoretic effects.  It is common in ${\cal N}=1$
gauge/gravity duality that the moduli space metric requires non-canonical K\"ahler potential in the field theory on the branes.  A different asymptotic behavior of the
K\"ahler potential could change the conclusion about the runaway, though any
asymptotically homogeneous K\"ahler potential will still yield a runaway -- either to large
or small field expectation values.  A minimum could only come from an inhomogeneity in the Kahler potential.

The outline of this short note is as follows.  In sect. 2, we
review the fact that anomalous $U(1)$ gauge fields are massive,
and that their D-term potentials should not be imposed
\refs{\DineXK, \DouglasSW}.  In sect. 3, we discuss the
$SU(3M)\times SU(2M)\times SU(M)$ quiver gauge theory of $M$
wrapped D5 branes on the complex cone over $F_1$.  The simplest
case is $M=1$, where the gauge group is $SU(3)\times SU(2)$, and
the matter content is similar to the 3-2 model of \AffleckXZ, but
with an extra pair of $SU(2)$ doublets and a particular
superpotential.  Though the 3-2 model of \AffleckXZ\ does
dynamically break supersymmetry with a stable groundstate, this
string inspired variant does not.  This  example also illustrates
that added vector-like matter can ruin supersymmetry breaking,
depending on what its tree-level superpotential couplings are.

\newsec{Comments on $SU(N)$ vs $U(N)$ in quiver gauge theories}

An issue that has been discussed by various authors is whether the
worldvolume quiver gauge theory of branes at singularities is
$\prod _i SU(N_i)$ or $\prod _i U(N_i)$.   The additional $U(1)$
factors in the latter case include a decoupled, diagonal, overall
$U(1)$ factor, under which no matter is charged.  We will not be
concerned with this $U(1)$ here.  The remaining $U(1)$ factors
have charged matter, and are hence IR free in four spacetime
dimensions.  When the string theory realization is via branes on a
local, non-compact Calabi-Yau, the low-energy gauge theory should
thus be considered as $\prod _i SU(N_i)$, because the $U(1)$
couplings vanish in the IR. These couplings can be taken to be
non-zero if the Calabi-Yau space is compact.

A distinct issue is the fact that the $U(1)$ factors of $\prod _i
U(N_i)$ are often anomalous, with e.g. non-zero $\Tr
SU(N_i)^2U(N_j)$ anomalies.  This is generic for chiral quiver
gauge theories, and a simple example is in D3 branes at a $ \Bbb
C^3/\Bbb Z_3$ orbifold singularity. As discussed in detail in
\DouglasSW, the worldvolume theory of the branes contains the
necessary coupling to implement the Green-Schwarz anomaly
cancellation mechanism as in \DineXK. The upshot is that anomalous
$U(1)$ gauge fields $A_\mu$ are Higgsed by coupling to scalars
$B$, through $(A_\mu -\partial _\mu B)^2$.

Since anomalous $U(1)$ gauge fields are massive, they are not
present in the low-energy effective field theory.  For this
reason, their D-term equations should not be imposed.
Equivalently, supersymmetry pairs $B$ with a field $\phi$, which
plays the role of the FI term for the anomalous $U(1)$ \DineXK.
The  D-term of the anomalous $U(1)$ gauge field can then always
relax to zero, by suitable expectation value $\ev{\phi}$, so it
does not constrain the low-energy fields.  This agrees with the
discussion in \FrancoZU, sect. 3.2.3.  See \DiaconescuPC\ for
a discussion in the context of compact Calabi-Yaus, where it is
suggested that some other dynamics could perhaps induce an
additional potential for the K\" ahler modulus $\phi$.

\newsec{The gauge theory}

The gauge theory of $M$ wrapped D5 branes on the complex cone over
$F_1$ is
 \eqn\gaugethy{
 \matrix{&SU(3M)&SU(2M)&SU(M)&[ SU(2) & U(1)_F & U(1)_R ] \cr \cr
 Q&{\bf 3M}&{\bf{\overline{2M}}}&{\bf 1}&{\bf 1}&1&-1\cr\cr
 \overline u&{\bf{\overline{3M}}}&{\bf 1}&{\bf M}&{\bf 2}&-1&0\cr\cr
 L&{\bf 1}&{\bf 2M}&{\bf{\overline{M}}}&{\bf 2}&0&3 \cr\cr
 L_3&{\bf 1}&{\bf 2M}&{\bf{\overline{M}}}&{\bf 1}&-3&-1,}}
where $SU(3M)\times SU(2M)\times SU(M)$ are the gauge symmetries,
and the groups in $\left[\cdot \right]$ are the global symmetries,
with $U(1)_R$ an R-symmetry. There is a tree-level superpotential
\eqn\wtree{W_{tree}=hQ\overline u_iL_j\epsilon ^{ij},} where we
here explicitly write the $SU(2)$ flavor indices $i,j=1,2$.

Note that we cannot extend $SU(3M)\rightarrow U(3M)$, because the
additional $U(1)_{3M}$ factor would be anomalous under both of the
other two groups, e.g. $\Tr U(1)_{3M}SU(2M)^2=3M$; similarly, we
cannot extend $SU(2M)\rightarrow U(2M)$ or $SU(M)\rightarrow
U(M)$, each of the additional $U(1)$ factors would be anomalous
under both of the other two gauge groups.

The couplings, and their charges under various symmetries (some
broken) are:
 \eqn\instch{ \matrix{&U(1)_Q&U(1)_{\bar
 u}&U(1)_{L}&U(1)_{L_3}&U(1)_F&U(1)_{R}\cr\cr \Lambda
 _{3M}^{7M}&2M&2M&0&0&0&0\cr\cr \Lambda
 _{2M}^{3M}&3M&0&2M&M&0&0\cr\cr \Lambda
 _M^{-3M}&0&6M&4M&2M&-12M&0\cr\cr h&-1&-1&-1&0&0&0.}}
The four symmetries $U(1)_{K=Q, \bar u, L, L_3}$ assign charge one
to $K=Q, \bar u, L, L_3$, and zero to all other fields; the
$U(1)_F$ and $U(1)_R$ charge assignments are as given in
\gaugethy\ ($U(1)_F \subset U(1)_Q\times U(1)_ {\bar u} \times
U(1)_{L}\times U(1)_{L_3}$). The $SU(M)$ group factor in
\gaugethy\ is IR free, as can be seen from the negative exponent
in its instanton factor in \instch. The $SU(2)$ and $U(1)_R$
global symmetries are preserved by the couplings. The symmetry
$U(1)_F$ arises as an accidental symmetry in the IR, as it is
broken only by the IR free group $SU(M)$.

Let us first consider the particular case of $M=1$, where the
gauge group \gaugethy\ is $SU(3)\times SU(2)$.  The matter content
of \gaugethy\ in this case is similar to that of the 3-2 model of
\AffleckXZ, but with an extra pair of $SU(2)$ doublets.

\subsec{The $SU(3)\times SU(2)$ theory ($M=1$), in the classical
limit.}

We initially consider the classical theory ($\Lambda_{2,3}\to 0$)
without the superpotential \wtree. In this case the three fields
$L$ and $L_3$ can be combined into an $SU(3)$ triplet
$L_{a=1,2,3}$.  The gauge invariant fields are
 \eqn\gaugeinvts{Z=\det_{fj} Q^f\overline u_j,
 \quad X_{ia}=Q\overline u_i L_a,
 \quad V^a={1\over 2}L_bL_c\epsilon ^{abc},}
where the color indices are suppressed (except for the $SU(2)$
color index $f$ in the expression for $Z$), and the flavor indices are
given.  These fields satisfy the classical constraints
 \eqn\constraints{ZV^a-{1\over 2} X_{ib}X_{jc}\epsilon
 ^{abc}\epsilon ^{ij}=0.}
The gauge group is completely broken for general expectation
values of these fields, and the complex dimension of the classical
moduli space of vacua is the number of fields left uneaten:
$6+6+6-3-8=7$.  This agrees with the description of the vacua in
terms of expectation values of the 10 fields \gaugeinvts, subject
to the 3 classical constraints \constraints.

Let us now consider the theory with added tree-level
superpotential
 \eqn\wzxt{W_{tree}=hX_{ij}\epsilon ^{ij},}
Since this interaction breaks the global $SU(3) \times SU(2)$
symmetry to $SU(2)$, we replace the index $a=1,2,3$ with $i=1,2$,
and will explicitly write the $a=3$ component, so the fields are
$Z$, $X_{ij}$, $X_i\equiv X_{i3}$, $V^i\equiv L_jL_3\epsilon
^{ij}$, and $V^{a=3}={1\over 2}L_iL_j\epsilon ^{ij}\equiv V$. The
superpotential \wzxt\ lifts the $Z$, $X_{ij}$, and $X_i$ classical
flat directions, but the $V$ and $V^i$ classical flat directions
remain unlifted.   These classical flat directions can be
parameterized in terms of the original microscopic fields, up to
gauge and flavor rotations, as
 \eqn\clasfa{L=(L_1,L_2)=\pmatrix{c&0\cr 0&d}\ , \qquad
 L_3=\pmatrix{\sqrt{|d|^2 - |c|^2} \cr 0},}
which give $V=cd$, $V^1=- d \sqrt{|d|^2 - |c|^2}$ and $V^2=0$,
with the other fields vanishing. Along these  unlifted flat
directions, the $SU(2)$ gauge group is Higgsed, and the $Q$ and
$\overline u_i$ matter fields get a mass from the superpotential
\wzxt.

There is thus a 3 complex dimensional classical moduli space of
vacua left unlifted by \wzxt.  The low-energy spectrum along this
classical moduli space is $SU(3)$ pure Yang-Mills, plus the 3
massless chiral superfields $V$ and $V^i$, in the ${\bf 1}$ and
${\bf 2}$ respectively of the global $SU(2)$ symmetry. Projecting
the classical K\" ahler potential, $K_{cl}=Q^\dagger Q+u^\dagger
u+L^\dagger L$ (all gauge and flavor indices are implicit and
summed over), on the unlifted classical moduli space of $V$ and
$V^i$ expectation values gives
 \eqn\clasK{K_{cl}( V,V^\dagger , V^i,(V^i)^\dagger
 )=2\sqrt{T}, \qquad T\equiv VV^\dagger+V^i(V^i)^\dagger=
 V^a(V^a)^\dagger.}
There is an accidental $SU(3)$ global symmetry, with the moduli
re-combined into $V^a$ in the ${\bf 3}$,  because the interactions
from the superpotential \wzxt, which had broken the global $SU(3)$
to $SU(2)$, do not affect the low-energy theory. Away from the
origin of the moduli space, the $SU(3)$ invariant $T\neq 0$, and
the $SU(3)$ symmetry is spontaneously broken to an $SU(2)$
subgroup.  The 6 real massless moduli from $V^a$ can then be
regarded the real modulus $T$, and the 5 Goldstone bosons from
$SU(3)/SU(2)$.

\subsec{The $SU(3)\times SU(2)$ quantum theory for $W_{tree}=0$.  }

Let us first consider the quantum theory, with $W_{tree}=0$, in
the limit $\Lambda _3\gg \Lambda _2$, where the $SU(2)$ dynamics
can be initially ignored.  The $SU(3)$ gauge group has $N_f=2$
flavors, so the dynamically generated superpotential \AffleckMK\
is
 \eqn\wi{W_{dyn}={\Lambda _3^7\over Z}.}

Let us now consider the opposite limit, $\Lambda _2\gg \Lambda
_3$. The $SU(2)$ gauge theory has $N_f=3$, so its low-energy
spectrum consists of the $SU(2)$ gauge invariant composites,
$V^a={1\over 2} L_bL_c\epsilon ^{abc}$, $QL_a$, and $Q^2$, with a
superpotential term \SeibergBZ. The $SU(3)$ gauge theory now has
$N_f=3$ flavors of fundamentals, $QL_a$, and anti-fundamentals,
$Q^2$ and $\overline u_i$, so its low-energy spectrum consists of
the gauge invariant fields with quantum deformed moduli space
constraint  \SeibergBZ.   In addition to the fields \gaugeinvts,
this yields the following fields, which are classically zero:
$Y_a={1\over 2}(Q^2)(QL_a)$, $B={1\over 6}(QL_a)(QL_b)(QL_c)
\epsilon ^{abc}$, with superpotential
 \eqn\wii{W_{dyn}=-{1\over \Lambda _2^3}\left(B-Y_aV^a
 \right)+C\left({1\over 2} Y_aX_{ib}X_{jc}\epsilon ^{ij}\epsilon
 ^{abc}- ZB-\Lambda _2^3\Lambda _3^7\right).}
$C$ is a Lagrange multiplier.  Integrating out the massive fields
$V^a$, $Y_a$, $B$, and $C$, we find $Y_a=0$, $B=-\Lambda
_2^3\Lambda _3^7/Z$, and we are left with the low-energy
superpotential \wi\ and the constraint \constraints.  It is also
seen from the symmetries \instch\ that the constraints
\constraints\ could not have been modified by quantum effects.

\subsec{The $SU(3)\times SU(2)$ quantum theory with
$W_{tree}=hX_{ij}\epsilon ^{ij}$, for $h\neq 0$.}

The full superpotential is given by adding $W_{tree}$ to the
dynamical superpotential \wi, with the  constraints \constraints\
imposed with Lagrange multipliers
 \eqn\wzx{W_{full}={\Lambda _3^7\over Z}+hX_{ij}\epsilon ^{ij}+
 \lambda_i (Z V^i +  X_j X_{kl} \epsilon^{jk}\epsilon^{il}) +
 \lambda(Z V - {1\over 2} X_{ij} X_{kl} \epsilon^{ik}\epsilon^{jl}).}
This leads to runaway of the field $V$.   Indeed, we can satisfy
the supersymmetric equations of motion for all other fields via
$Z=\left({ \Lambda_3^{14}\over  V h^2}\right)^{1\over 3}$,
$X_{ij}=\epsilon _{ij}\left({V\Lambda _3^7\over h}\right)^{1\over
3}$, $\lambda_i=0, \lambda = \left({ h^4 \over  V \Lambda_3^7 }
\right)^{1\over 3}$, $X_i= -\epsilon_{ij} V^j \left({ \Lambda
_3^7\over h V^2 } \right)^{1\over 3}$. The low-energy spectrum
consists of the fields $V$ and $V^i$, with the dynamical
superpotential
 \eqn\wlow{W_{low}=3(h^2V\Lambda _3^7)^{1/3}.}
Using  \gaugethy\ and \instch, this can be seen to be compatible
with all the symmetries.

Here is another way to quickly reproduce the superpotential \wlow:
the low-energy theory on the classical moduli space includes an
unbroken $SU(3)$ Yang-Mills theory, with no light matter.  The two
$SU(3)$ flavors get a mass $\sim h\sqrt{V}$ from $W_{tree}$, so
the scale of the low-energy $SU(3)$ Yang-Mills theory is given by
the matching relation  \eqn\wlowscale{\Lambda
_{3,low}^9=h^2V\Lambda _3^7.} The superpotential \wlow\ arises
from gaugino condensation in the $SU(3)$ Yang-Mills theory.

The superpotential \wlow\ lifts the classical moduli $V$ and
$V^i$, with potential
 \eqn\veffvc{V_{eff}=K^{VV^\dagger}\left| h^2\Lambda _3^7\right|
 ^{2/3}(VV^\dagger)^{-2/3},}
where $K^{VV^\dagger}$ is a component of the inverse K\" ahler
metric, computed on the 3 complex dimensional moduli space of $V$
and $V^i$.   In the limit of large $V$, we can use the classical
K\" ahler potential \clasK.  The quantum contributions of the
$SU(2)$ gauge group become negligible, because it is broken at a
high scale.  The quantum contributions of the $SU(3)$ gauge group
also become negligible for large $V$ or $V^i$, even though it
remains unbroken.   It is intuitively reasonable that the $SU(3)$
dynamics does not affect $K_{eff}$ for large $V, V^i$, as the only
matter fields that couple $SU(3)$ to $L,L_3$ are the fields $Q$,
which decouple as they become very massive.  In this limit, the
low-energy theory has the accidental $SU(3)$ global symmetry
discussed following \clasK, and the effective K\" ahler  potential
must be of the form
 \eqn\Keff{K_{eff}\approx K_{cl}=2\sqrt{T}=2 \sqrt{ VV^\dagger
 +V^iV^i{}^\dagger }}
i.e.\ we can use the classical K\" ahler potential \clasK\ in
\veffvc.  We conclude that
 \eqn\veffv{V_{eff}\approx |h^2 \Lambda_3^7|^{2/3} {2V^\dagger V
 +(V^i)^\dagger V^i \over \sqrt{V^\dagger V +(V^i)^\dagger V^i}}
 (V V^\dagger)^{-2/3}.}
In the parameterization \clasfa\ of the classical moduli space,
the potential \veffv\ is
 \eqn\potf{V_{eff} \approx |h^2 \Lambda_3^7|^{2\over 3}
 {|c|^2 +|d|^2 \over |cd|^{4/3}}.}
For fixed $|V|$, the potential is minimized by $|V^i|=0$, and the
remaining potential is
 \eqn\veffcc{V_{eff}\approx 2|h^2
 \Lambda_3^7|^{2/3}(VV^\dagger)^{-1/6}.}
There is thus a $V\rightarrow \infty$ runaway, and in that limit
supersymmetry is restored.

The above runaway direction is present not only on the moduli
space of the classical theory with nonzero $h$, but also on the
larger classical moduli space of the theory with $h=0$.
Furthermore, it is not isolated on a separated branch. This can
also be seen by working in terms of the D-flat microscopic fields,
writing the F-term potential as
 \eqn\vfis{V_{eff}=\left|
 {\Lambda _3^7\over Z^2}Q\overline u\overline u+h\overline
 uL\right| ^2+\left| {\Lambda _3^7\over Z^2}Q^2\overline u
 +hQL\right| ^2+\left| hQ\overline u\right| ^2 }
(we suppressed the color and flavor indices which are summed
over).  Extremizing in the fields that can have supersymmetric minima,
the only light fields are the classical moduli $V$ and $V^i$, with
low-energy potential \veffv.  There are no other branches.

To summarize, we have found that the effective potential has
runaway behavior and no metastable SUSY-breaking minimum, at least
in the regime of large $\ev{V}$, where the K\"ahler potential is
approximately canonical.  The dynamical  supersymmetry breaking of
the original $SU(3)\times SU(2)$ model of \AffleckUZ\ would be
recovered if we  could add a mass term $W_{mass}=mL_2L_3$ to the
tree-level superpotential.  But, in the context of this brane
system, such a mass  term is forbidden by the $SU(2)$ global
symmetry; this global symmetry is ensured by the isometry of the
space $dP_1$.  There is a mass term which  respects the $SU(2)$
global symmetry, $W_{mass'}=-{m\over 2}L_iL_j \epsilon ^{ij}=-
mV$, but adding that mass term leads to a supersymmetric ground
state with $\langle V \rangle=h \Lambda_3^{7/2}/ m^{3/2}$.  As $m
\to 0$ this states moves to infinity.

\subsec{The generalization to $M>1$ wrapped branes.}

The gauge theory \gaugethy\ with superpotential $W_{tree}$ \wtree\
has  runaway direction:
 \eqn\lmvevs{(L_1, L_2)=\pmatrix{c{\bf 1}_{M\times M}&{\bf 0}
 _{M\times M}\cr
 {\bf 0}_{M\times M}& c{\bf 1}_{M\times M}},}
with ${\bf 1}_{M\times M}$ an $M\times M$ unit matrix, and ${\bf
0}_{M\times M}$ a vanishing $M\times M$ matrix. The gauge
invariant light field along this direction is $V\equiv
\det_{2M\times 2M}(L_1, L_2)=c^{2M}$, and the group is Higgsed as
$SU(3M)\times SU(2M)\times SU(M)\rightarrow SU(3M)\times SU(M)'$,
where $SU(M)'$ is a diagonal subgroup of $SU(2M)\times SU(M)$.
The $SU(3M)$ group has no massless flavors, as  the fields $Q$ and
$\overline u_i$ get a mass along this direction, because of
$W_{tree}$ \wtree.  The $SU(M')$ have two massless adjoints,
coming from the fields $L_3$. The fields $L_3$ also yield  two
massless singlets, corresponding to the gauge invariants $V_i=\det
_{2M\times 2M}(L_i, L_3)$.

The dynamical scales of the low-energy theory are given by the
matching relations
 \eqn\lamlowm{\Lambda _{3M, low}^{9M}=h^{2M}V\Lambda
 _{3M}^{7M}, \qquad \Lambda _{M, low}^{M}=(\Lambda
 _{2M}^{3M})^2\Lambda _M^{-3M}/V,}
where the exponents in the latter scale relation are because of
the index of the embedding of $SU(M)'\subset SU(2M)\times SU(M)$.
The effective superpotential along the runaway direction comes
from gaugino condensation in the low-energy $SU(3M)$ Yang-Mills
theory (the $SU(M)'$ theory has too much massless matter to
dynamically generate a superpotential):
 \eqn\wmrunaway{W_{low}=3M\left(h^{2M}V\Lambda
 _{3M}^{7M}\right)^{1/3M}.}
The canonical K\"ahler potential for $V$ is $K_{can}\sim
(VV^\dagger)^{1/2M}$, so the potential for large $V$ is
$V_{eff}\sim \left|(h^{2M}\Lambda ^{7M})^2V^{-1}\right|^{1/3M}$,
which again has a runaway $\ev{V}\rightarrow \infty$.

\vskip 0.8cm

\noindent {\bf Acknowledgments:}

We would like to thank I.~Bena, I.~Klebanov, J.~Maldacena and
especially D.~Berenstein, S.~Kachru, D.~Shih for useful
discussions. The research of NS is supported in part by DOE grant
DE-FG02-90ER40542. The research of KI is supported in part by UCSD
grant DOE-FG03-97ER40546 and by the IAS Einstein Fund. KI would
like to thank the IAS for their hospitality and support on his
sabbatical visit.  NS would like to thank the hospitality of The
Institute for Advanced Studies of The Hebrew University in
Jerusalem where this work was completed.

\listrefs

\end